\documentclass[reprint,pra, twocolumn, amsmath,amssymb,
 aps,superscriptaddress,]{revtex4-2}

\usepackage[utf8]{inputenc}
\usepackage[english]{babel}
\usepackage[a4paper,margin=2.0cm]{geometry}


\usepackage{indentfirst}
\usepackage{graphicx}
\usepackage{booktabs}
\usepackage{float}
\usepackage{siunitx}
\usepackage{lipsum}
\usepackage{listings}
\usepackage{amsmath}
\usepackage{amsfonts}
\usepackage{tasks}
\usepackage{amssymb}
\usepackage{wasysym}
\usepackage[T1]{fontenc}

\usepackage{tikz}
\usepackage{comment}
\usepackage{mathtools}
\usepackage{extpfeil}
\usepackage{xcolor}
\usepackage{hyperref}
\usepackage{url}
\usepackage{natbib}
\usetikzlibrary{external}
\newcommand{\ie}{\textit{i.e.}}

\newcommand{\ic}{$\mathcal{IC}$}

\newcommand{\ba}[1]{\begin{align} #1 \end{align}}

\newtheorem{lemma}{Lemma}
\newtheorem{result}{Result}
\begin{document}

\title{Security of device-independent quantum key distribution via monogamy relations from multipartite information causality}
\author{Lucas Pollyceno}
\email{lucas.da.silva.pollyceno@ug.edu.pl}
\affiliation{Instituto de Física “Gleb Wataghin”, Universidade Estadual de Campinas, 13083-859, Campinas, Brazil}
\affiliation{International Centre for Theory of Quantum Technologies (ICTQT), University of Gda{\'n}sk, 80-308 Gda\'nsk, Poland}

\author{Anubhav Chaturvedi}
\email{anubhav.chaturvedi@pg.edu.pl}
\affiliation{International Centre for Theory of Quantum Technologies (ICTQT), University of Gda{\'n}sk, 80-308 Gda\'nsk, Poland}
\affiliation{Faculty of Applied Physics and Mathematics, Gda{\'n}sk University of Technology, Gabriela Narutowicza 11/12, 80-233 Gda{\'n}sk, Poland}

\author{Chithra Raj}
\affiliation{International Centre for Theory of Quantum Technologies (ICTQT), University of Gda{\'n}sk, 80-308 Gda\'nsk, Poland}
\author{Pedro R. Dieguez}
\affiliation{International Centre for Theory of Quantum Technologies (ICTQT), University of Gda{\'n}sk, 80-308 Gda\'nsk, Poland}
\author{Marcin Paw{\l}owski}
\affiliation{International Centre for Theory of Quantum Technologies (ICTQT), University of Gda{\'n}sk, 80-308 Gda\'nsk, Poland}

\begin{abstract}

The \emph{information causality} principle ($\mathcal{IC}$) was originally proposed as a promising means to understand the extent of quantum nonlocality without invoking the full mathematical formalism of quantum theory. Beyond the foundational significance, the problem of bounding nonlocal correlations by reasonable physical principles has meaningful practical consequences, particularly for device-independent (DI) cryptographic security. In this work, we advance in this direction, demonstrating that the $\mathcal{IC}$ is enough to ensure DI security on quantum key distribution (QKD) protocols. Security is proven for a range of theoretically quantum-attainable parameters against individual attacks by a potentially post-quantum eavesdropper. This result follows as a consequence of a strong form of monogamy of Bell’s inequality violations, which we have proven to be implied by the recently proposed multipartite formulation for $\mathcal{IC}$. Additionally, we demonstrated that the original bipartite formulation of $\mathcal{IC}$ fails to imply monogamy relations and hence, ensure security of DIQKD, thus stressing the necessity of the multipartite framework.

\end{abstract}
\maketitle

\noindent\emph{Introduction:---} Quantum theory predicts strong correlations between spatially separated observers, which defy local-causal explanations ~\cite{Bell1964}. Apart from their foundational significance, nonlocal quantum correlations power several classically inconceivable information processing tasks ~\cite{PhysRevLett.67.661, PhysRevLett.91.187904,vandam2005implausible, PhysRevLett.95.010503, PhysRevLett.97.120405, cui2024monogamynonlocalgames}. In particular, quantum theory only allows a peculiarly restricted amount of nonlocality~\cite{Tsirelson_lim}, which can be estimated using the quantum formalism. However, the abstract mathematical formalism of quantum theory does not offer any insights into the underlying reasons for nature's restraint on nonlocality. Consequently, numerous efforts have recently been made to recover the set of nonlocal quantum correlations from well-motivated physical principles \cite{NontrivialCC, ML, OL, D’Ariano_Chiribella_Perinotti_2017}. 

The most prominent among such proposals, is the physical principle of \emph{information causality} ($\mathcal{IC}$) \cite{IC}, which states that the amount of randomly accessible data cannot exceed the capacity, $C$, of a classical communication channel even when it is assisted by nonlocal correlations \cite{IC, Allcock_2009, Yang_2012}. As originally suggested in 2009 \cite{IC}, $\mathcal{IC}$ extends the idea of NS principle to communicating scenarios -- whereas $C \rightarrow 0$, $\mathcal{IC}$ reduces to no-signalling. Thus, $\mathcal{IC}$ forbids stronger-than-quantum nonlocal correlations to a significant extent ~\cite{IC, Allcock_2009, Yang_2012}. The most outstanding feat of $\mathcal{IC}$ is the recovery of the \emph{Tsirelson's} bound on the maximum quantum violation of the Clauser-Horne-Shimony-Holt (CHSH) inequality. While quantum theory satisfies $\mathcal{IC}$, it remains unclear whether all stronger-than-quantum nonlocal correlations necessarily violate $\mathcal{IC}$.

The main challenge in deriving bounds on nonlocal correlations with $\mathcal{IC}$ springs from the inherent dependence of the principle on the communication protocol. A recent contribution \cite{ICnoisy}, made significant headway towards reducing the complexity and broadening the application of $\mathcal{IC}$ by replacing the poorly-scaling concatenation procedure with \emph{noisy channels}. However, another significant short-coming of the initial formulation $\mathcal{IC}$ is that it is bipartite, and \emph{any} operational principle which seeks to recover the quantum set of nonlocal correlations must be \emph{multipartite} \cite{Yang_2012, PhysRevLett.107.210403}. To address this issue, $\mathcal{IC}$ was recently reformulated to apply to multipartite communication scenarios and shown to forbid nontrivial \emph{stronger-than-quantum} multipartite nonlocal correlations \cite{multiIC}.

Although the precise boundary of quantum nonlocality remains unexplained in terms of fundamental physical principles, such as information causality ($\mathcal{IC}$), its deep connection with cryptographic tasks constitutes one of the most striking practical consequences of quantum theory \cite{PhysRevLett.97.120405, PhysRevLett.98.230501}. In particular, the quantum key distribution (QKD) framework ensures security based solely on the laws of quantum mechanics without relying on any assumptions about the computational power of a potential eavesdropper. The device-independent quantum key distribution (DIQKD) framework extends this idea further by making no assumptions about the physical description of the devices used by the parties, instead grounding security in fundamental operational principles \cite{PhysRevLett.97.120405}. Under this paradigm, when parties are constrained by the no-signaling principle—prohibiting communication faster than the speed of light—security proofs may remain valid even in the face of a hypothetical post-quantum breakthrough that grants an adversary access to supra-quantum resources. On the other hand, a broad class of supra-quantum no-signaling correlations has been consistently shown to be likely non-physical, as they would lead to implausible consequences —for instance, making the communication complexity problem trivial\cite{vandam2005implausible} or even violating the $\mathcal{IC}$ principle \cite{IC}. Consequently, this fact raises the fundamental question of whether device-independent security proofs can be established based on alternative physical principles.

In this work, we take a significant step toward demonstrating that the $\mathcal{IC}$ principle ensures the security of DIQKD protocols. In particular, by employing a refined multipartite informational-theoretic approach, we have shown that the so-called CHSH protocol \cite{PhysRevLett.97.120405} is safe when parties respect the multipartite operational form for the $\mathcal{IC}$ principle. Notably, the security holds for a range of quantum-attainable violations of the CHSH inequality. This result follows directly from a nontrivial monogamy relation for CHSH inequalities, which we derive from the multipartite formulation of $\mathcal{IC}$. In particular, when two parties observe the maximal violation of the CHSH inequality, multipartite $\mathcal{IC}$ strictly prohibits any correlation with a third party, thereby fully recovering the quantum monogamy of Bell inequality violations. For non-maximal violations, multipartite $\mathcal{IC}$ imposes tighter than no-signaling bounds, effectively limiting nonlocal correlations with a third party and enhancing the security of DIQKD against an adversary constrained by $\mathcal{IC}$. Additionally, we demonstrate that previous bipartite formulations of $\mathcal{IC}$ fail to impose the same monogamy constraints on CHSH inequalities, highlighting the necessity of the multipartite framework for $\mathcal{IC}$.
  
 \textit{Preliminaries:---} Let us begin by revisiting the essential preliminaries for the CHSH Bell experiment entailing two distant parties, Alice $(\mathcal{A})$ and Bob $(\mathcal{B})$. In each round of the experiment, Alice and Bob independently randomly choose their inputs $x,y\in\{0,1\}$, and retrieve outputs $a,b\in\{0,1\}$, respectively. Their results produce a joint probability distribution $p(a,b|x,y)$, referred to as a \emph{correlation}. A correlation $p(a,b|x,y)$ is deemed \emph{nonlocal} if and only if it violates the CHSH inequality, defined as,
\begin{align}\label{eq:chsh}
\beta(\mathcal{A},\mathcal{B}) = \frac{1}{4} \sum_{x,y} p(a\oplus b = xy | x, y) \le \frac{3}{4}.
\end{align}
Here, $\oplus$ represents the sum modulo 2. The value of the CHSH functional, $\beta(\mathcal{A},\mathcal{B})$, not only determines the presence of nonlocal correlations but also quantifies their strength. Specifically, the completely uncorrelated case yields $1/2$, whereas the maximum value allowed by local causality is capped at $3/4$. According to quantum theory, the CHSH inequality can be violated up to $\beta_Q = \frac{1}{2}(1+\frac{1}{\sqrt{2}}) \approx 0.8535$, a characteristic limit known as the Tsirelson's bound \cite{cirel1980quantum}. Correlations which satisfy the no-signaling condition can attain an even higher violation of CHSH inequality up to $\beta_{NS} = 1$. In general, correlations which violate the CHSH inequality exhibit several nonclassical features, the most notable and cryptographically significant being their \emph{monogamy}. 

\textit{Monogamy of nonlocality} restricts the extent to which nonlocal correlations can be shared between multiple spatially separated parties. Specifically, let us consider a tripartite Bell scenario including $\mathcal{A},\mathcal{B}$, and an additional party $\mathcal{E}$ with an input $z\in\{0,1\}$ and an output $e\in\{0,1\}$. Then monogamy of nonlocality implies that if the two parties $\mathcal{A},\mathcal{B}$ observe nonlocal correlations, such that  $\beta(\mathcal{A},\mathcal{B})>3/4$, then the strength of their correlations with $\mathcal{E}$, as measured by the value of the CHSH functional $\beta(\mathcal{B},\mathcal{E})$ (or  $\beta(\mathcal{A},\mathcal{E})$) remains limited. This notion is captured by means of a \emph{monogamy relation} of the generic form,  
\begin{align}\label{eq:monogamy_T}
\beta(\mathcal{B},\mathcal{E}) \le f_T^M \left(\beta(\mathcal{A},\mathcal{B})\right),
\end{align}
where $f_T^M : [1/2 , 1] \mapsto [0,1]$ is a function specifying the characteristic monogamy relation of given nonlocal theory $T$. Monogamy relations of the form \eqref{eq:monogamy_T} are cryptographically significant as they can be used to derive criteria for ensuring the security of DIQKD against adversaries restricted by the nonlocal theory $T$ \cite{PhysRevA.82.032313}. In particular, for the DIQKD protocol based on the CHSH set-up considered in \cite{PhysRevA.82.032313,PhysRevLett.97.120405}, the sufficient condition from ensuring security against individual attacks translates to \cite{CommentOnMarcin,ReplyByMarcin},
\begin{align}\label{eq:sec_mono}
    h(\beta(\mathcal{A},\mathcal{B})) < 3 - 4 f_T^M ( \beta(\mathcal{A},\mathcal{B}) ),
\end{align} 
where $h(p) = -p\log p - (1-p)\log (1-p)$ is the Shannon's binary entropy.  
The condition \eqref{eq:sec_mono} implies threshold values of $\beta(\mathcal{A},\mathcal{B})$ required for the security, which can be obtained by substituting $f_T^M$ in \eqref{eq:sec_mono}, with the monogamy relation \eqref{eq:monogamy_T}, for any given nonlocal theory $T$.

For instance, correlations satisfying the no-signaling conditions obey the following \emph{linear} monogamy relation\cite{Monogamy_NS},
\begin{align}\label{eq:monogamy_ns}
 \beta(\mathcal{B},\mathcal{E}) \le \frac{3}{2}-\beta(\mathcal{A},\mathcal{B}).
\end{align}
Notice that when $\beta(\mathcal{A},\mathcal{B})=\beta_{NS}=1$, the no-signaling condition implies that $\mathcal{B}$ (and $\mathcal{A}$) must be completely uncorrelated with the third party $\mathcal{E}$, such that, $\beta(\mathcal{B},\mathcal{E})=1/2$. The threshold value of $\beta(\mathcal{A},\mathcal{B})$ for secure DIQKD with no-signaling monogamy \eqref{eq:monogamy_ns} turns out to be $\approx 0.881$, which is not realizable with quantum theory \cite{ReplyByMarcin}. Of particular relevance, quantum theory features a tighter than no signaling, characteristic \emph{quadratic} monogamy relation \cite{toner2006monogamy},
\begin{align}\label{eq:monogamy_q}
\left(\beta(\mathcal{A},\mathcal{B})-\frac{1}{2}\right)^2 + \left(\beta(\mathcal{B},\mathcal{E})-\frac{1}{2}\right)^2 \le \frac{1}{8}.
\end{align}
Similar to the no-signaling case, when $\mathcal{A},\mathcal{B}$ observe $\beta(\mathcal{A},\mathcal{B})=\beta_{Q}=\frac{1}{2}(1+\frac{1}{\sqrt{2}})$, then $\beta(\mathcal{B},\mathcal{E})$ must be $1/2$. In this case, the threshold value of $\beta(\mathcal{A},\mathcal{B})$ for secure DIQKD turns out to be $\approx0.841$ \eqref{eq:sec_mono}.  While \eqref{eq:monogamy_q} holds for quantum theory, we are interested in whether a non-trivial, i.e., tighter than \eqref{eq:monogamy_ns}, monogamy relation of the form \eqref{eq:monogamy_T} can be derived via a physical principle, without invoking the abstract Hilbert space formalism. Towards this end, we now present the physical principle of information causality.

\textit{Information causality ($\mathcal{IC}$):---} The principle of $\mathcal{IC}$ is typically formulated by means of a bipartite communication task, namely, the $(n\mapsto m)$ \emph{random access code} (RAC), wherein the parties utilise a nonlocal correlation assisted by a classical communication channel of bounded capacity \cite{PhysRevA.96.022125}. Specifically, the sender ($\mathcal{A}$) receives a randomly sampled bit string $\mathbf{x}\equiv(X_1,\ldots,X_{n})\in\{0,1\}^n$ of length $n$. $\mathcal{A}$ then encodes $\mathbf{x}$ onto a classical message $M$ of $m$ bits with $m<n$. The message $M$ is then transmitted through a noisy classical channel with capacity $C\le m$, such that $\mathcal{B}$ gets $M'$. $\mathcal{B}$ then decodes the message $M'$ to produce a guess $G_i$ about a randomly selected bit of $X_i$ of $\mathcal{A}$, where $i\in\{1,\ldots,n\}$. In this set-up, the principle of $\mathcal{IC}$ states that
\textit{the total potential information $\mathcal{B}$ can gain about the $\mathcal{A}$'s bit string $\mathbf{x}$ cannot exceed the capacity $C$ of the classical communication channel}, i.e., 
\ba{\label{eq:firstICnoisy}\sum_{i=1}^{n} I(X_i:G_i) \le C,}
where $I(X_i:G_i)$ denotes Shannon's mutual information between $\mathcal{A}$'s $i$-th input $X_i$ and the corresponding $\mathcal{B}$'s guess $G_i$, $\sum_{i=1}^{n} I(X_i:G_i)$ is the total potentially accessible information.  \footnote{We note here that in \eqref{eq:firstICnoisy} we are considering the most recent definition proposed in \cite{ICnoisy}, which is a generalization of the original proposal \cite{IC}.}. 

Quantum theory satisfies $\mathcal{IC}$, while stronger than quantum nonlocal correlations violate $\mathcal{IC}$, and hence, are ruled out by the principle. Specifically, let us consider the $(2\mapsto 1)$ RAC, and let $\mathcal{A},\mathcal{B}$ share a no-signaling PR-box correlation \cite{PRbox}, defined as: $p(a,b|x,y)=\frac{1}{2}\delta_{a\oplus b,xy}$. The parties then perform the van Dam protocol \cite{vandam2005implausible}, wherein $\mathcal{A}$ inputs $x=X_1\oplus X_2$ into her part of the PR-box, and transmits the message $M=a\oplus X_1$. Let the classical communication be noiseless such that $C=1$ and $M'=M$. $\mathcal{B}$ upon receiving $M$ randomly inputs $y=i-1$ into his PR-box to produce the outcome $G_i=M\oplus b$. Since the PR-box satisfies $a\oplus b = xy$, $G_i=X_i$ and $I(X_1:G_1)=I(X_2:G_2)=C=1$, thereby violating the $\mathcal{IC}$ criterion \eqref{eq:firstICnoisy}. Hence, the PR-box is ruled out by $\mathcal{IC}$. Moreover, any nonlocal correlation $p(a,b|x,y)$ which violates the CHSH inequality beyond the Tsirelson's bound $\beta(\mathcal{A},\mathcal{B})>\beta_{Q}=\frac{1}{2}(1+\frac{1}{\sqrt{2}})$ is ruled out by $\mathcal{IC}$, for some $C\in [0,1]$ \cite{IC,ICnoisy}.

Moreover, we can derive an even more general criterion for $\mathcal{IC}$ by considering the causal structure associated with the $(n\mapsto m)$ RAC and using the technique described in \cite{Chaves_2015} to compute $\mathcal{IC}$-like \footnote{satisfied by both classical and quantum theory.} information theoretic constraints for arbitrary causal structures. Specifically, the method \cite{Chaves_2015} returns the following generalized criterion for $\mathcal{IC}$ which takes into non-uniform priors and arbitrary decoding protocols, 
\begin{equation}
\begin{aligned}\label{eq:recentIC} \sum_{i=1}^{n}I(X_i : G_i,M') + \sum_{i=2}^{n}I(X_1 : X_i| G_i,M') \\ \le C + \sum_{i=2}^{n}H(X_i) - H(X_1,...,X_{n}),\end{aligned}\end{equation}
where $H(V)$ denotes the Shannon's entropy of the argument random variable $V$. Up to this point, we have invoked the notion of $\mathcal{IC}$ in association with a bipartite communication task. We now present a refined form of the recently proposed multipartite criterion for $\mathcal{IC}$ \cite{multiIC}.
\begin{figure}[t!]
    \centering
    \includegraphics[width=1\columnwidth]{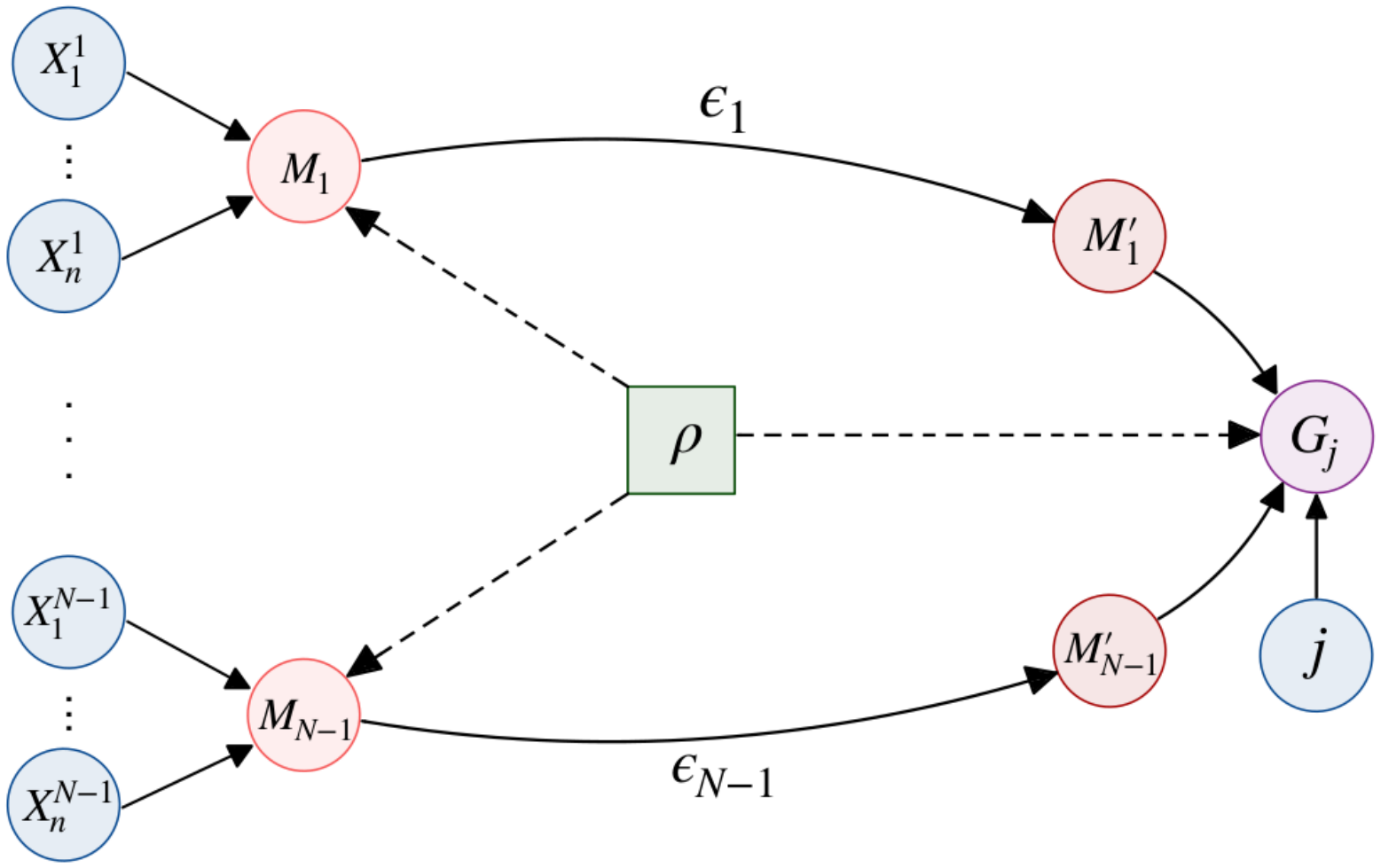}
    \caption{The graphic depicts the causal structure represented as a Directed Acyclic Graph (DAG) associated with the communication task for the multipartite $\mathcal{IC}$ criterion \eqref{eq:multipartite_IC}, entailing $N-1$ senders and a receiver. The parties have access to a pre-shared entangled quantum state $\rho$ (green square). The senders $\{\mathcal{A}_k\}^{N-1}_{k=1}$ receive inputs $\{\{X^k_j\}^n_{j=1}\}^{N-1}_{k= 1}$ (blue disks), and transmit messages $\{M_k\}^{N-1}_{k=1}$ (pink disks) through binary-symmetric noisy classical channels with parameters $\{\epsilon_k\}^{N}_{k=1}$, to the receiver $\mathcal{B}$, respectively. Upon receiving the $N-1$ potentially noisy messages $\{M'_{k}\}^{N-1}_{k=1}$, the receiver computes guess $G_j$ (purple disk) about a joint function $f_j(\{X^k_j\}^{N-1}_{k= 1})$, based on a randomly selected input $j\in \{1,\ldots n\}$ (green disk).
    \label{NewScenarioIC}}
\end{figure}

\textit{Multipartite information causality:---} Consider a communication task entailing $N$ spatially separated parties wherein $N-1$ senders $\{\mathcal{A}_k\}^{N-1}_{k=1}$ transmit information to a receiver $\mathcal{B}$. Similar to the $(n\mapsto m)$ RAC, each sender $\mathcal{A}_k$ receives an $n$-bit string $\mathbf{x}^k = (X_1^k, X_2^k, \cdots, X_n^k)$ as input which is encoded on to a $m<n$-bit classical message $M_k$ and transmitted to $\mathcal{B}$ via a potentially noisy communication channel with capacity $C_k<m$. $\mathcal{B}$ in turn gets the noisy messages $\{M_k'\}$, and based on randomly chosen input $j\in\{1,\ldots,n\}$ he produces a guess $G_j$ about a function of the form $f_j (X_j^1, X_j^2, \cdots, X_j^{N-1})$. The directed acyclic graph representing the causal structure associated with this communication task in presented in Figure \ref{NewScenarioIC}. Observe that in contrast to bipartite $(n\mapsto m)$ RAC wherein $\mathcal{B}$ attempts to guess a randomly chosen bit of $\mathcal{A}$, this task is multipartite, as $\mathcal{B}$ guesses a joint function $f_j$ of the $j$th input bits $\{X_j^k\}^{N-1}_{k=1}$ of the $N-1$ senders $\{\mathcal{A}_k\}^{N-1}_{k=1}$. Consequently, the bipartite criteria of $\mathcal{IC}$ presented above \eqref{eq:firstICnoisy} and \eqref{eq:recentIC} fail to discard post-quantum correlations in this set-up. Instead, let us consider the following multipartite criterion for $\mathcal{IC}$, 
\begin{multline}\label{eq:multipartite_IC}
   \sum_{k,i} I(X_i^k : X_i^1, \dots,X_i^{k-1}, X_i^{k+1}, \dots, X_i^{N-1}, \mathbf{M'}, G_i) \\ \le  
    \sum_{k=1}^{N-1} C_k + \sum_{i=1}^n I(X_{i+1}^k, \dots, X_n^k :  X_i^k).
\end{multline}
where $\mathbf{M'} = (M_1' , M_2', \cdots, M_{N-1}')$ denotes the tuple of the messages reaching the receiver through the $N$ potentially noisy classical channels. Note that the criterion \eqref{eq:multipartite_IC} generalizes the one presented in \cite{multiIC} by accounting for the noisy received message on the receiver's knowledge about the initial data. In the supplementary material, we show that theories—such as classical and quantum mechanics—that satisfy the information-theoretic axioms of $\mathcal{IC}$ also fulfill this condition and further explain why this refinement is essential for deriving monogamy relations.

In contrast to bipartite criteria \eqref{eq:firstICnoisy}, \eqref{eq:recentIC}, the multipartite criterion \eqref{eq:multipartite_IC} for $\mathcal{IC}$ forbids stronger than quantum no-signaling correlations in this set-up (Figure \ref{NewScenarioIC}). Specifically, let us consider the simplest tripartite version of the multipartite communication task presented above, with $N=3$ and $n=2$. Let the two senders $\mathcal{A},\mathcal{E}$ and the receiver $\mathcal{B}$ share a tripartite no-signaling correlation $p(a,b,e|x,y,z)$ satisfying $a\oplus b \oplus e = xy \oplus zy$.  $\mathcal{A},\mathcal{E}$ input uncorrelated bits $x=X_0^1 \oplus X_1^1$, $z=X_0^2 \oplus X_1^2$, and transmit the message $M_1 = a \oplus W_0^1$ and $M_2 = e \oplus X_0^2$ via noiseless communication channels such that $C_1=C_2=1$, respectively. Upon receiving the messages $\{M_k\}^2_{k=1}$, $\mathcal{B}$ then inputs $y=j-1$ and produces the guess $G_{j} = M_1 \oplus M_2 \oplus b$. Consequently, we have that $I(X_i^k : X_i^{3-k}, M_1, M_2 , G_i ) = C_k = 1$, and $I(X_{i+1}^k, \dots, X_n^k :  X_i^k)=0$ for all $k,i\in\{1,2\}$, which violates the multipartite $\mathcal{IC}$ criterion \eqref{eq:multipartite_IC}, thereby, ruling out the no-signaling correlation. We are now ready to examine the secrecy of cryptographic protocols when parties are limited by the $\mathcal{IC}$ principle.

\begin{figure}[t!]
    \centering    \includegraphics[width=1\columnwidth]{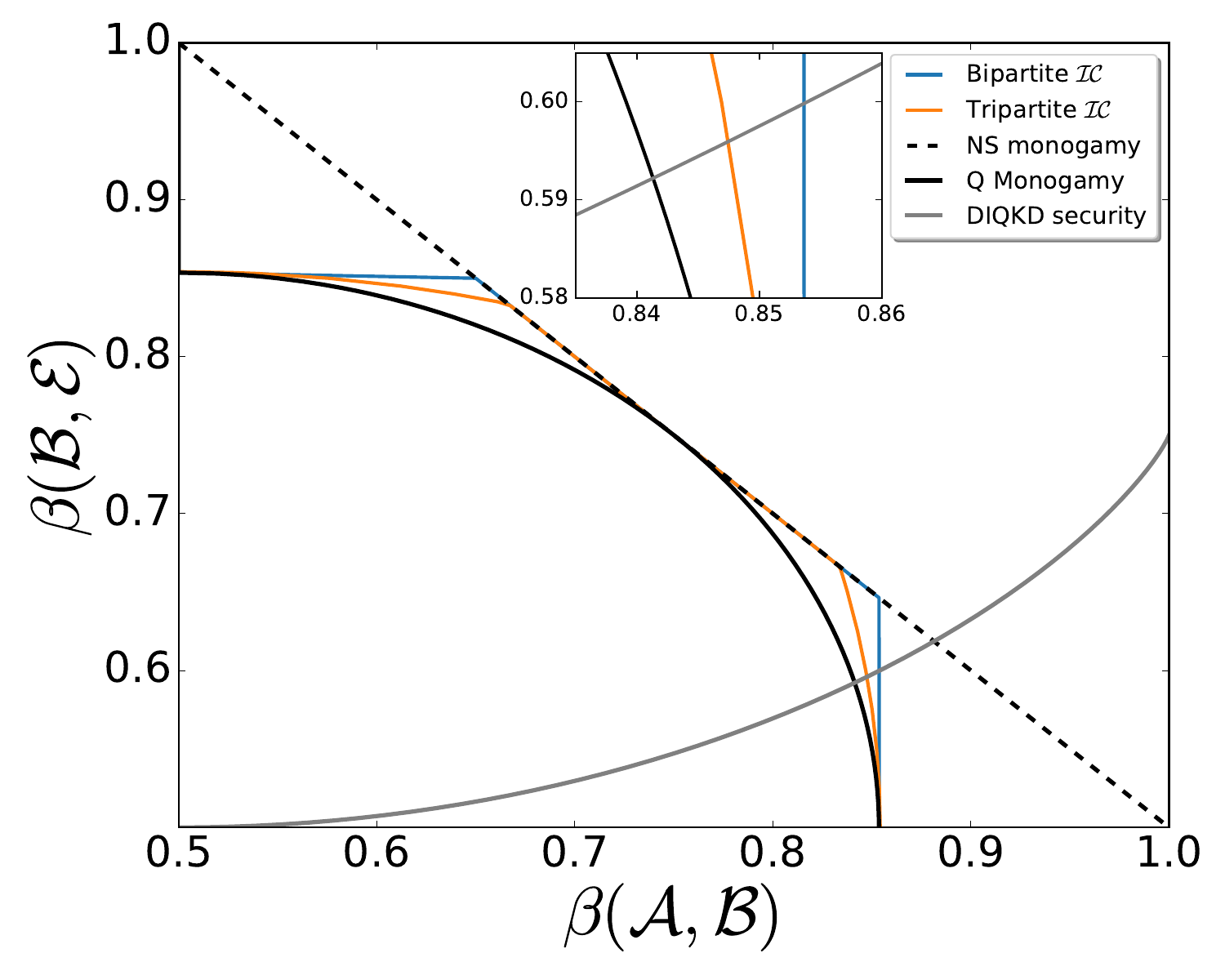}
    \caption{
    A plot of the maximum value of the CHSH functional $\beta(\mathcal{B},\mathcal{E})$ implied by the monogamy relations (of the form \eqref{eq:monogamy_T}) considered in this work, against the CHSH functional $\beta(\mathcal{A},\mathcal{B})\in[1/2,1]$. The dashed and solid black lines represent the monogamy relations implied by the no-signaling condition \eqref{eq:monogamy_ns} and quantum theory \eqref{eq:monogamy_q}, respectively. The solid blue line represents monogamy relation implied by the bipartite $\mathcal{IC}$ criteria \eqref{eq:firstICnoisy},\eqref{eq:recentIC} taking into account all possible wiring of the form \eqref{eq:wiring}. The solid orange line represents the monogamy relation implied by the tripartite $\mathcal{IC}$ criterion \eqref{eq:tripartite_IC}. Finally, the solid gray line exhibits the security condition  \eqref{eq:sec_mono} for DIQKD. Notice that, in contrast to the bipartite criterion, the tripartite $\mathcal{IC}$ criterion implies a non-trivial monogamy relation for $\beta (\mathcal{A}, \mathcal{B}) \in [0.8333,\beta_Q=\frac{1}{2}(1+\frac{1}{\sqrt{2}})]$, and ensures security of DIQKD for $\beta (\mathcal{A}, \mathcal{B}) \in [0.8471,\beta_Q]$.}
    \label{fig:slice}
\end{figure}

\emph{Optimal slice:---}  As introduced in the preamble, the secrecy of DIQKD protocols has close relations with monogamy of Bell inequalities violation, as follows from \eqref{eq:sec_mono}. Consequently, to assess the $\mathcal{IC}$ principle's ability to ensure security, we investigate its connection with monogamy relations in the form \eqref{eq:monogamy_T}. In this regard, the problem reads as finding the maximum value of $\beta(\mathcal{B},\mathcal{E})$ given $\beta(\mathcal{A},\mathcal{B})$ over all tripartite no-signaling correlations which satisfy the respective \emph{non-linear} and \emph{protocol-dependent} $\mathcal{IC}$ criteria. Thus, such procedure shall be performed for each bipartite \eqref{eq:firstICnoisy},\eqref{eq:recentIC} and multipartite \eqref{eq:multipartite_IC} criteria. Since the convex polytope of tripartite no-signaling correlations has $53856$ extremal points \cite{322paper}, these optimization problems are particularly complex. We now present a useful Lemma that significantly reduces this complexity and allows us to restrict to a two-parameter slice of the tripartite no-signaling polytope, 
\begin{lemma} \label{lemma:optSlice}
To find the maximum CHSH value between $\mathcal{B}$ and $\mathcal{E}$, $\beta(\mathcal{B},\mathcal{E})$, permitted by information causality, when $\mathcal{A}$ and $\mathcal{B}$ witness a CHSH value, $\beta(\mathcal{A},\mathcal{B})$, it suffices to consider tripartite no-signaling correlations $p(a,b,e|x,y,z)$ of the form,
\begin{multline}\label{eq:slice}
    p(a,b,e|x,y,z) = \alpha \frac{1}{4}\delta_{a\oplus b, x y} + \gamma \frac{1}{4}\delta_{e\oplus b, z y} + \\ +(1-\alpha -\gamma)1/8,
\end{multline}
where $\alpha,\gamma\in[0,1]$, and $\alpha+\gamma\leq 1$.
\end{lemma}
The proof follows from the \emph{data-processing} inequality and has been deferred to the supplementary material for brevity. Notice that, for any point on the optimal slice specified by $\{\alpha,\gamma\}$ \eqref{eq:slice} the values CHSH functionals are specified as $\beta(\mathcal{A},\mathcal{B})=\frac{1+\alpha}{2}$ and  $\beta(\mathcal{B},\mathcal{E})=\frac{1+\gamma}{2}$. Hence, our problem reduces to finding the maximum $\gamma$ given $\alpha$, such that the correlation \eqref{eq:slice} satisfies the $\mathcal{IC}$ criteria \eqref{eq:firstICnoisy},\eqref{eq:recentIC},\eqref{eq:multipartite_IC}. These problems can now be efficiently tackled, \emph{up to machine precision}, as we describe below \footnote{codes containing the numerical solutions are available online at \cite{github}}. We plot the resultant curves in Figure \ref{fig:slice}. 
\begin{figure}[t!]
    \centering
    \includegraphics[width=1.05\columnwidth]{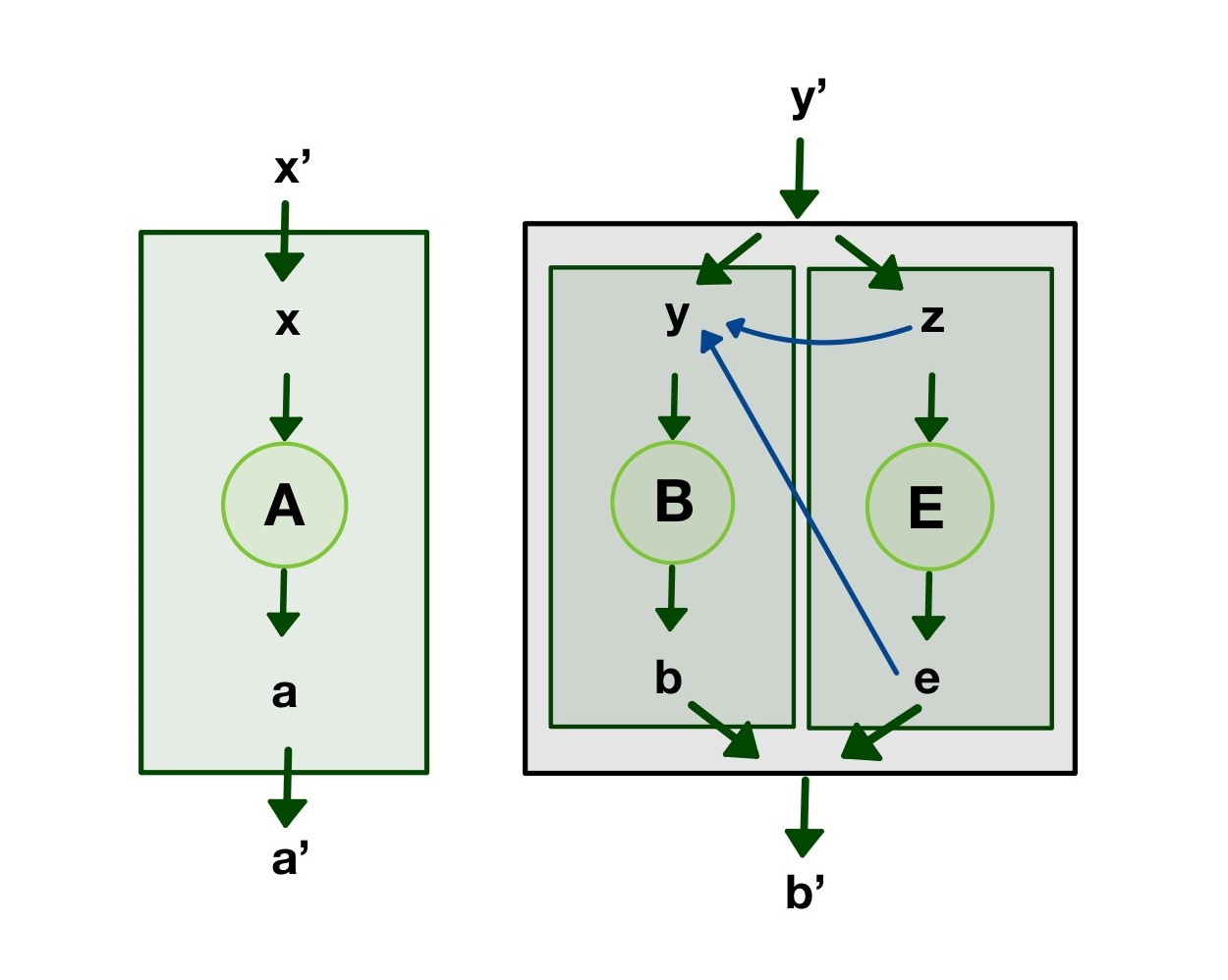}
    \caption{Wiring procedure that takes tripartite correlations $p(a,b,e|x,y,z)$ and produces a bipartite effective one $p_{eff}(a',b'|x',y')$.}
    \label{fig:wiring}
\end{figure}

\emph{No monogamy from bipartite $\mathcal{IC}$:---}Before presenting our main findings, we emphasize the necessity of a multipartite framework for $\mathcal{IC}$ in the context of monogamy relations. Specifically, we demonstrated in Figure \ref{fig:slice} that both the original \eqref{eq:firstICnoisy} and the generalized \eqref{eq:recentIC} bipartite formulations of $\mathcal{IC}$ fail to establish non-trivial monogamy relations for CHSH inequalities, such as \eqref{eq:monogamy_T}. In this regard, we emphasize that the criteria \eqref{eq:firstICnoisy} and \eqref{eq:recentIC} are essentially bipartite. Consequently, a tripartite no-signaling correlation $p(a,b,e|x,y,z)$ must be \emph{locally} post-processed into an effectively bipartite correlation $\tilde{p}(a',b'|x',y')$ in order to be tested against them. Moreover, we need only consider deterministic post-processing schemes, referred to as \emph{wirings}. A wiring is completely specified by the choice of the bipartition, for instance $\mathcal{A}$ and $\mathcal{B}'\equiv (\mathcal{B},\mathcal{E})$, and the functions, 
\begin{align}\label{eq:wiring}
    x &= F_1 (x'), &a' =& F_2 (a), &&\\  \nonumber\\
    y &= F_3 (y',z,e), &z =& F_4 (y'),\: &b'& = F_5 (b, e),\nonumber
\end{align}
where, $F_i : \{0,1\}^n \mapsto \{0,1\}, \; \forall i \in \{1,2,3,4,5\}$. 
An example of such a wiring is depicted in Figure \ref{fig:wiring}. 
We note here that the grouped parties $\mathcal{B},\mathcal{E}$ may signal to each other. Consequently, a tripartite correlation $p(a,b,e|x,y,z)$ violates the bipartite criterion \eqref{eq:recentIC} for $\mathcal{IC}$, if there exists some wiring that produces an effectively bipartite correlation $\tilde{p}(a',b'|x',y')$ which violates it. In fact, such an approach has previously been used to study $\mathcal{IC}$ in multipartite Bell scenarios \cite{Yang_2012, PhysRevLett.107.210403, Xiang}. In particular, \cite{Hsu, Emily1, Emily2} claimed to have derived the quantum monogamy relation \eqref{eq:monogamy_q} through the bipartite $\mathcal{IC}$ criteria \eqref{eq:firstICnoisy} by employing a wiring of the form \eqref{eq:wiring}. Contrary to these claims, we find that neither the original \eqref{eq:firstICnoisy} nor the generalized $\mathcal{IC}$ \eqref{eq:recentIC} criteria imply
stronger-than-no-signaling monogamy relations, for \emph{any} wiring of the form \eqref{eq:wiring}.

Specifically, for all tripartite correlations $p(a,b,e|x,y,z)$ of the form \eqref{eq:slice} with $\alpha,\gamma \in [0,1]$, we consider all possible wirings of the form \eqref{eq:wiring}, to retrieve the effectively bipartite correlations $\tilde{p}(a',b'|x',y')$. Then, for each such $\tilde{p}(a',b'|x',y')$, we employ the aforementioned protocol for the $(2\mapsto 1)$ RAC, with a binary symmetric noisy communication channel which flips the message bit $M$ with a probability $p(M'=M\oplus 1|M)=\epsilon \in [0,1/2)$ for $M\in\{0,1\}$. Paralleling the observation in \cite{ICnoisy}, we find that the tightest bounds on the maximum of value $\gamma$ given $\alpha\in[0,1]$ are recovered as $\epsilon \rightarrow 1/2$. We plot the resultant monogamy relation in Figure \ref{fig:slice}, which is commented on in the main text. In this case, the bipartite $\mathcal{IC}$ criteria \eqref{eq:firstICnoisy},\eqref{eq:recentIC} retrieve the Tsirelson's bounds, such that $\beta(\mathcal{B},\mathcal{E})\leq \beta_Q= \frac{1}{2}(1+\frac{1}{\sqrt{2}})$ for $\beta(\mathcal{A},\mathcal{B})\in [1/2,\frac{1}{2}(2-\frac{1}{\sqrt{2}})]$. However, for $\beta(\mathcal{A},\mathcal{B})\in[\frac{1}{2}(2-\frac{1}{\sqrt{2}}),\beta_Q]$ the monogamy relation implied by these criteria coincides with the no-signaling monogamy relation \eqref{eq:monogamy_ns} such that, $\beta(\mathcal{B},\mathcal{E})\leq \frac{3}{2}-\beta(\mathcal{A},\mathcal{B})$. Consequently, we conclude that the original \eqref{eq:firstICnoisy} and generalized \eqref{eq:recentIC} bipartite criteria for $\mathcal{IC}$ fail to yield non-trivial monogamy relations beyond no-signaling. In particular, when $\mathcal{A},\mathcal{B}$ observe the Tsirelson's bound $\beta(\mathcal{A},\mathcal{B})=\beta_Q$, the bipartite criteria \eqref{eq:firstICnoisy},\eqref{eq:recentIC} play no role for monogamy relations. In this case, in fact, while quantum monogamy from \eqref{eq:monogamy_q} implies $\beta(\mathcal{B},\mathcal{E})\leq \frac{1}{2}$, the bipartite criteria yield $\beta(\mathcal{B},\mathcal{E})\leq\frac{1}{2}(2-\frac{1}{\sqrt{2}})$. As we shall demonstrate in this section, addressing this limitation requires the more refined approach for $\mathcal{IC}$ that inherently accounts for multipartite scenarios, as formalized in \eqref{eq:multipartite_IC}.

\emph{Monogamy from multipartite $\mathcal{IC}$:---} Building on this refined framework, we analyze tripartite correlations of the form \eqref{eq:slice} on a specific communication task. More precisely, for each tripartite correlation $p(a,b,e|x,y,z)$ of the form \eqref{eq:slice}, we consider the protocol described above for the simplest non-trivial multipartite communication task with $(N=3,n=2)$. Furthermore, we assume independent binary symmetric noisy classical channels between $\mathcal{A},\mathcal{B}$ and $\mathcal{E},\mathcal{B}$,  respectively, which flip the input with probability $p(M'_1=M_1\oplus 1|M_1)=\epsilon_1$ and $p(M'_2=M_2\oplus 1|M_1)=\epsilon_2$, such that multipartite $\mathcal{IC}$ criterion \eqref{eq:multipartite_IC} translates to,
\begin{align}\label{eq:tripartite_IC}
    \sum^2_{k=1}\sum^2_{j=1}I(X_j^k : X_j^{3-k}, M_1', M_2' , G_j )
    \le 2 - \sum^2_{k=1}h(1-\epsilon_k).
\end{align}

In contrast to the bipartite $\mathcal{IC}$ criteria \eqref{eq:firstICnoisy},\eqref{eq:recentIC}, we find that there exists correlations of the form \eqref{eq:slice}, which satisfy the no-signalling monogamy relation \eqref{eq:monogamy_ns}, but violate the tripartite $\mathcal{IC}$ criterion \eqref{eq:tripartite_IC} for some $\epsilon_1,\epsilon_2\in [0,1/2)$. In other words, the tripartite $\mathcal{IC}$ criterion \eqref{eq:tripartite_IC} implies a non-trivial monogamy relation of the form \eqref{eq:chsh}, which we plot in Figure \ref{fig:slice}. Most significantly, when $\mathcal{A},\mathcal{B}$ observe the Tsirelson's bound $\beta(\mathcal{A},\mathcal{{B}})=\frac{1}{2}(1+\frac{1}{\sqrt{2}})$, the criterion \eqref{eq:tripartite_IC} is violated for all $\beta(\mathcal{B},\mathcal{E})>1/2$, up to machine precision. This forms our first result, 
\begin{result}
        When $\mathcal{A}$ and $\mathcal{B}$ witness the maximum quantum violation of the CHSH inequality such that $\beta(\mathcal{A},\mathcal{{B}})=\beta_Q$, tripartite information causality \eqref{eq:tripartite_IC} implies $\mathcal{B}$ must be completely uncorrelated with any third party $\mathcal{E}$, such that the CHSH value between $\mathcal{B}$ and $\mathcal{E}$ must be $\beta(\mathcal{B},\mathcal{E})= 1/2$, thereby recovering the quantum monogamy \eqref{eq:monogamy_q}.
 \end{result}
Moreover, the criterion \eqref{eq:tripartite_IC} retrieves tighter than no-signaling monogamy relations for $\beta(\mathcal{A},\mathcal{B}) \in [0.8333, \beta_Q]$. In the supplementary material, we show how the refinement provided in Eq.~\eqref{eq:multipartite_IC} plays a crucial role in the derivation of this result. We now demonstrate that the monogamy relation implied by \eqref{eq:tripartite_IC}, although weaker than the quantum monogamy relation \eqref{eq:monogamy_q}, is strong enough to enhance the information-theoretic security of DIQKD protocols.

\emph{Secure DIQKD from $\mathcal{IC}$:---} Recall that the generic monogamy relations of the form \eqref{eq:monogamy_T} can be used to derive the condition \eqref{eq:sec_mono} for ensuring security of DIQKD protocol based on the CHSH scenario against individual attack of an eavesdropper constrained by a nonlocal theory $T$  \cite{PhysRevA.82.032313,PhysRevLett.97.120405,CommentOnMarcin,ReplyByMarcin}. In Figure \ref{fig:slice}, we plot the condition \eqref{eq:sec_mono} for ensuring security of DIQKD protocol. Observe that neither the no-signaling condition \eqref{eq:monogamy_ns} nor the bipartite criteria \eqref{eq:firstICnoisy},\eqref{eq:recentIC} yield security for realizable quantum correlations $\beta(\mathcal{A},\mathcal{B})\leq \beta_Q$. In contrast, we find that in the range  $\beta(\mathcal{A},\mathcal{B})\in[0.8471,\beta_Q]$, the monogamy relation implied by the tripartite $\mathcal{IC}$ criterion \eqref{eq:tripartite_IC} satisfies the criterion \eqref{eq:sec_mono}, thereby, ensuring the security of DIQKD, which forms our main result,

\begin{result}
    Tripartite information causality \eqref{eq:tripartite_IC} ensures security of DIQKD whenever $\mathcal{A}$ and $\mathcal{B}$ witness the CHSH value in the realizable range of quantum correlations, $\beta(\mathcal{A},\mathcal{B})\in[0.8471,\beta_Q]$.
\end{result}

The results establish a strong connection between a largely foundational field and practical applications. Notably, the security based on $\mathcal{IC}$ surpasses that based on no-signaling monogamy relations, and holds within the range $\beta(\mathcal{A},\mathcal{B}) \in [0.8471, \beta_Q]$, which are attainable by means of quantum experiments. This raises the question of whether $\mathcal{IC}$-based security proofs can consistently outperform those grounded solely in the no-signaling principle. In this context, more refined security criteria and amplification protocols could significantly enhance these findings. Furthermore, since $\mathcal{IC}$ imposes tighter constraints than no-signaling, our results suggest a potential breakthrough in device-independent (DI) proofs within the quantum key distribution (QKD) paradigm.

To summarize, theoretical security proofs in device-independent quantum key distribution (DIQKD) are often based solely on the no-signaling principle without assuming any specific underlying physical theory describing the apparatus. However, strong evidence—such as that from the information causality ($\mathcal{IC}$) principle—suggests that a wide range of no-signaling correlations have no correspondence in nature, as they lead to implausible consequences. This observation raises the fundamental question of whether theoretical security proofs can also be drawn based on stronger-than-no-signaling reasonable principles.
In consonance with this concern, here we demonstrated that when parties are assumed to be limited by the $\mathcal{IC}$ principle, security holds for a range of physically achievable parameters. A central open question regarding $\mathcal{IC}$ is whether it fully characterizes the set of quantum correlations. Nevertheless, our results reveal that, even in the negative case, if $\mathcal{IC}$ can be trusted as nature's physical principle, secure cryptographic keys can be extracted against potential supra-quantum attacks. This follows directly from the nontrivial monogamy properties of Bell inequality violations derived from the multipartite formulation of $\mathcal{IC}$. Notably, the multipartite $\mathcal{IC}$ principle recovers the strong quantum monogamy of maximal CHSH inequality violations without relying on the Hilbert space formalism.
Our findings highlight the necessity of the multipartite framework by demonstrating that bipartite formulations of $\mathcal{IC}$ alone do not impose monogamy constraints on CHSH inequalities. Given the inherent protocol-dependent to any formulation of the $\mathcal{IC}$ criteria, it remains open whether this gap can be closed through alternative protocols, as well as including non-binary-symmetric noisy classical channels in the analysis.
In conclusion, building upon significant developments in the DIQKD literature regarding the no-signaling principle, our results provide a foundation for exploring the implications of stricter device-independent operational principles in practical applications, such as quantum key distribution (QKD) and quantum random number generation (QRNG). Promising natural directions for future research include analyzing security under general eavesdropping scenarios—such as collective or coherent attacks \cite{PhysRevLett.98.230501}—exploring amplification protocols \cite{PhysRevA.74.042339}, and extending our approach to more general cryptographic settings, including protocols based on chained Bell inequalities \cite{Acin_2006}.

We thank Rafael Rabelo for fruitful discussions and valuable suggestions. This study was financed in part by the Coordenação de Aperfeiçoamento de Pessoal de Nível Superior - Brasil (CAPES) - Finance Code 001 - and by the Brazilian National Council for Scientific and Technological Development (CNPq). This work is partially carried out under IRA Programme, project no. FENG.02.01-IP.05-0006/23, financed by the FENG program 2021-2027, Priority FENG.02, Measure FENG.02.01., with the support of the FNP. This work was partially supported by the Foundation for Polish Science (IRAP project, ICTQT, contract No. MAB/218/5, co-financed by EU within the Smart Growth Operational Programme). AC acknowledges financial support by NCN grant SONATINA 6 (contract No. UMO-2022/44/C/ST2/00081). C.R acknowledges support from the Narodowe Centrum Nauki (NCN) (SHENG project, contract No. 2018/30/Q/ST2/00625) and partially by the National Centre for Research and Development (QuantERA project, contract no. QUANTERA/2/2020). P.R.D acknowledges support from the NCN Poland, ChistEra-2023/05/Y/ST2/00005 under the project Modern Device Independent Cryptography (MoDIC).

\clearpage
\appendix
\onecolumngrid

\section*{Supplementary Material}

In this appendix, we provide proofs for the updated version of multipartite information causality ($\mathcal{IC}$) introduced in the equation \eqref{eq:multipartite_IC},  Lemma \ref{lemma:optSlice}, and no-monogamy relations from bipartite formulated criteria for $\mathcal{IC}$.

\subsection*{Proof of multipartite $\mathcal{IC}$ criterion 
\eqref{eq:multipartite_IC}}

The main ingredients of this proof are the axioms of the original $\mathcal{IC}$ formulation, namely, the chain rule of the mutual information and the data processing inequality, which are respectively expressed as,
\begin{align}
    &I(A:B|C) = I(A:B,C) - I(A:C); \label{chain}\\
    &I(A:B') \le I(A:B), \quad \text{where} \quad B \longrightarrow B'.\label{eq:dataprocessing}
\end{align}
The latter states that any local manipulation of data can only decay information
where $I(A : B)$ is an abstract theory independent mutual information between systems $A,B$, and $B'$ is obtained from $B$ via a local transformation. In addition to these two requirements, our proof also relies on another theory-independent axiom of $\mathcal{IC}$ (see \cite{IC}), known as \textit{consistency}, which asserts that the mutual information $I(A : B)$ must reduce to Shannon's mutual information when both subsystems $A$ and $B$ are classical.
With the multipartite communication scenario depicted in Figure \ref{NewScenarioIC}, let us consider the mutual information quantity of the bit-string of the $k$-th part, $\mathbf{x}^k$, with the data and remaining parties, local resources of the receiver, $c$, and all received messages $\{M'_k\}^{N-1}_{k=1}$ denoted as $\mathbf{M'}$, \ie,   $I(\mathbf{x}^k : \mathbf{x}^1, \cdots, \mathbf{x}^{k-1}, \mathbf{x}^{k+1}, \cdots, \mathbf{x}^{N-1}, \mathbf{M'}, c )$. This quantity provides a measure of the entire network's knowledge about the data set of part $k$. In this case, for brevity, we may consider the relation obtained in Ref. \cite{multiIC} (see Eq. (B8)), which states that when both chain rule and data processing conditions hold, we may straightly write,
\begin{multline}
    I(\mathbf{x}^k : \mathbf{x}^1, \cdots, \mathbf{x}^{k-1}, \mathbf{x}^{k+1}, \cdots, \mathbf{x}^{N-1}, \mathbf{M'}, c ) \\ \ge \sum_{i=1}^n I(X_i^k : X_i^1, X_i^2, \cdots,X_i^{k-1}, X_i^{k+1}, \cdots, X_i^{N-1}, \mathbf{M'}, c) - \sum_{i=1}^n I(X_{i+1}^k, \cdots, X_n^k :  X_i^k). \label{ap1:one}
\end{multline}
Notice that \eqref{ap1:one} is independent of the relation that the involved random variables may have among them.

Given that $\mathbf{M'}$ are classical variables, we use the data processing inequality \eqref{eq:dataprocessing} to refine the above inequality as, 
\begin{multline}
    I(\mathbf{x}^k : \mathbf{x}^1, \cdots, \mathbf{x}^{k-1}, \mathbf{x}^{k+1}, \cdots, \mathbf{x}^{N-1}, \mathbf{M'}, c ) \\ \ge \sum_{i=1}^n I(X_i^k : X_i^1, X_i^2, \cdots,X_i^{k-1}, X_i^{k+1}, \cdots, X_i^{N-1}, \mathbf{M'}, G_i)- \sum_{i=1}^n I(X_{i+1}^k, \cdots, X_n^k :  X_i^k). \label{ap1:two}
\end{multline}

Next, we derive an upper bound on $I(\mathbf{x}^k : \mathbf{x}^1, \cdots, \mathbf{x}^{k-1}, \mathbf{x}^{k+1}, \cdots, \mathbf{x}^{N-1}, \mathbf{M'}, c )$ by decomposing $\mathbf{M'}$ into $M_1', \cdots, M_k', \cdots, M_{N-1}'$ and applying the chain rule, to obtain, 
\begin{align}\label{ap1:three}
    I(\mathbf{x}^k : \mathbf{x}^1, \cdots, \mathbf{x}^{k-1}, \mathbf{x}^{k+1}, \cdots, &\mathbf{x}^{N-1}, M_1', \cdots, M_k', \cdots, M_{N-1}', c ) \\ = &I\:(\mathbf{x}^k : M_k' | \mathbf{x}^1, \cdots, \mathbf{x}^{k-1}, \mathbf{x}^{k+1}, \cdots, \mathbf{x}^{N-1},M_1', \cdots, M_{k-1}', M_{k+1}', \cdots,M_{N-1}', c)\nonumber\\&+ I(\mathbf{x}^k : \mathbf{x}^1, \cdots, \mathbf{x}^{k-1}, \mathbf{x}^{k+1}, \cdots, \mathbf{x}^{N-1},M_1', \cdots, M_{k-1}', M_{k+1}', \cdots,M_{N-1}', c). \nonumber
\end{align}

The second term on the right-hand side vanishes due to the no-signaling assumption. Applying the chain rule to the remaining term and using the non-negativity of mutual information $I(A:B)\ge 0$, we get
\begin{multline}\label{ap1:four}
    I(\mathbf{x}^k : \mathbf{x}^1, \cdots, \mathbf{x}^{k-1}, \mathbf{x}^{k+1}, \cdots, \mathbf{x}^{N-1}, M_1', \cdots, M_k', \cdots, M_{N-1}', c )\\ \le I(M_k' : \mathbf{x}^1, \cdots, \mathbf{x}^{k}, \cdots, \mathbf{x}^{N-1},M_1', \cdots, M_{k-1}', M_{k+1}', \cdots,M_{N-1}', c).
\end{multline}

Applying the data processing inequality \eqref{eq:dataprocessing} again, we know that including $M_k$ in the right-hand side can only increase the mutual information,
\begin{multline}\label{ap1:five}
    I(\mathbf{x}^k : \mathbf{x}^1, \cdots, \mathbf{x}^{k-1}, \mathbf{x}^{k+1}, \cdots, \mathbf{x}^{N-1}, M_1', \cdots, M_k', \cdots, M_{N-1}', c )\\ \le I(M_k' : \mathbf{x}^1, \cdots, \mathbf{x}^{k}, \cdots, \mathbf{x}^{N-1},M_1', \cdots, M_{k-1}', M_{k+1}', \cdots,M_{N-1}', c, M_k).
\end{multline}

From the causal structure depicted in Figure \ref{NewScenarioIC}, it is clear that $M_k$ shields $M_k '$ from all other variables $\mathbf{V}$, such that $\;I(M_{k}' : \mathbf{V}| M_{k}) = I(M_{k}' : \mathbf{V}, M_{k}) - I(M_{k}' : M_{k})= 0$. Thus, we simplify the inequality \eqref{ap1:five} as, 
\begin{align}\label{ap1:six}
    I(\mathbf{x}^k : \mathbf{x}^1, \cdots, \mathbf{x}^{k-1}, \mathbf{x}^{k+1}, \cdots, \mathbf{x}^{N-1}, M_1', \cdots, M_k', \cdots, M_{N-1}', c )\\ \le I(M_k' : M_k) = C_k. \nonumber
\end{align}

Finally, by combining \eqref{ap1:two} and \eqref{ap1:six}, and summing over all $k$, we recover the $\mathcal{IC}$ criterion \eqref{eq:multipartite_IC} presented in the main text,  
\begin{align}\label{ap1:seven}
   \sum_{k=1}^{N-1}\sum_{i=1}^n I(X_i^k : X_i^1, \dots,X_i^{k-1}, X_i^{k+1}, \dots, X_i^{N-1}, \mathbf{M'}, G_i) \le  
    \sum_{k=1}^{N-1} C_k + \sum_{k=1}^{N-1}\sum_{i=1}^n I(X_{i+1}^k, \dots, X_n^k :  X_i^k).
\end{align}

We stress that keeping this particular form for the multipartite $\mathcal{IC}$ criterion becomes crucial for the results presented in the main text. In fact, the previous formulation presented in \cite{multiIC} does not explicitly account for the received message on the measure of the receiver knowledge about the initial data (\ie, $I(X_i^k : X_i^1, \dots,X_i^{k-1}, X_i^{k+1}, \dots, X_i^{N-1}, G_i)$). It becomes clear by analyzing the same scenario considered for the results of the main text in the noiseless case (three parties and the same protocol employed by the parties). In the extreme case where parties share a PR-box, $p(a,b,e|x,y,z) = \delta_{a\oplus b  = xy}/4$, the receiver guessing might be written as $G_{i} =  X_i^1\oplus M_2$, which leads the mutual information terms present in \eqref{eq:multipartite_IC} to $I(X_i^k : X_i^{3-k}, M_1, M_2 , G_i ) = 1$ , while for the previous criterion we will have $I(X_i^k : X_i^{3-k}, G_i ) = 0$. The same can be obtained considering the other extreme case where $p(a,b,e|x,y,z) = \delta_{e\oplus b  = zy}/4$, and $G_{i} =  X_i^2\oplus M_1$. Consequently, the previous multipartite criterion of \cite{multiIC} in terms of $I(X_i^k : X_i^{3-k}, G_i )$ witness no the violation of $\mathcal{IC}$ for the optimal slice \eqref{eq:slice}, considering the same particular encoding and decoding protocols. Thus, being unable to recover the monogamy results presented in Fig.\ref{fig:slice}.

\subsection*{Proof of Lemma 1}

In this section, we prove Lemma \ref{lemma:optSlice}. In particular, we show that the slice defined in Eq.\eqref{eq:slice} is optimal for deriving monogamy relations of the form \eqref{eq:monogamy_T}, based on the bipartite and multipartite information causality (\ic) criteria given in Eqs. \eqref{eq:firstICnoisy}, \eqref{eq:recentIC}, and \eqref{eq:tripartite_IC}, within the tripartite Bell scenario involving parties $\mathcal{A}$, $\mathcal{B}$, and $\mathcal{E}$, each with binary inputs $x,y,z\in{0,1}$ and binary outputs $a,b,e\in{0,1}$.

First of all, we start by recalling the \textit{depolarization} procedure described in \cite{GeneralProperties, PhysRevA.74.042339}, which states that any bipartite distribution $\mathbb{P}_{\mathcal{A}, \mathcal{B}}$ can be transformed into the isotropic distribution using local operations and shared randomness (LOSR), keeping the CHSH value. That is,
\begin{subequations}\label{ap:lemma1}
\begin{align}
   \mathbb{P}_{\mathcal{A} \mathcal{B}} \quad\quad &\xmapsto{\text{LOSR}_{\mathcal{A} \mathcal{B}}} \quad \mathbb{P}_{\mathcal{A} \mathcal{B}}^{\text{iso}(\alpha)} = \alpha \text{PR}_{\mathcal{A}\mathcal{B}} +(1-\alpha)\text{W}_{\mathcal{A}\mathcal{B}},\\ \nonumber\\\beta(\mathcal{A},\mathcal{B})_{\mathbb{P}_{\mathcal{A} \mathcal{B}}} &\quad\:=\quad \quad\beta(\mathcal{A},\mathcal{B})_{\mathbb{P}_{\mathcal{A} \mathcal{B}}^{\text{iso}(\alpha)}} = \frac{1+\alpha}{2},
\end{align}
\end{subequations}
where $\alpha \in[0,1]$, $\text{PR}_{\mathcal{A}\mathcal{B}}$ is the PR-box, achieving the algebraic maximum value of the CHSH expression, $\beta(\mathcal{A},\mathcal{B})$ (\ie, $p_{\text{PR}_{\mathcal{A}\mathcal{B}}}(a,b|xy) = (1/2)\delta_{a\oplus b,xy} $), and $\text{W}_{\mathcal{A}\mathcal{B}}$ denotes the \textit{white noise} distribution (\ie, $p_{\text{W}_{\mathcal{A}\mathcal{B}}} (a,b|xy) = 1/4$). 

The depolarization procedure extends naturally to tripartite distributions. That is, any distribution $\mathbb{P}_{\mathcal{A} \mathcal{B} \mathcal{E}}$ can be transformed under $\text{LOSR}_{\mathcal{A} \mathcal{B}}$ applied to parties $\mathcal{A}$ and $\mathcal{B}$ into an isotropic tripartite distribution $\mathbb{P}_{\mathcal{A} \mathcal{B}\mathcal{E}}^{\text{iso}}$, while preserving the CHSH value $\beta(\mathcal{A},\mathcal{B})$. That is, $\mathbb{P}_{\mathcal{A} \mathcal{B} \mathcal{E}} \xmapsto{\text{LOSR}_{\mathcal{A} \mathcal{B}}} \mathbb{P}_{\mathcal{A} \mathcal{B}\mathcal{E}}^{\text{iso}}$, such that $\beta(\mathcal{A},\mathcal{B})_{\mathbb{P}_{\mathcal{A} \mathcal{B} \mathcal{E}}} =\beta(\mathcal{A},\mathcal{B})_ {\mathbb{P}_{\mathcal{A} \mathcal{B} \mathcal{E}}^{\text{iso}}}$, and:
\begin{align}\label{ap:lemma2}
    \sum_{\mathcal{E}} \mathbb{P}_{\mathcal{A} \mathcal{B}\mathcal{E}}^{\text{iso}} = \alpha \text{PR}_{\mathcal{A}\mathcal{B}} +(1-\alpha)\text{W}_{\mathcal{A}\mathcal{B}}.
\end{align}
Here, $\beta(\mathcal{A},\mathcal{B})_{\mathbb{P}_{\mathcal{A} \mathcal{B} \mathcal{E}}}$ refers to the CHSH value obtained by marginalizing the tripartite distribution over the party $\mathcal{E}$. Since the PR-box is an extremal point in the tripartite Bell scenario with binary inputs and outputs \cite{322paper}, the isotropic distribution $\mathbb{P}_{\mathcal{A} \mathcal{B}\mathcal{E}}^{\text{iso}}$ must be a convex combination of the PR-box and another distribution $\mathbb{P}_{\mathcal{A} \mathcal{B}\mathcal{E}}'$ satisfying the condition \eqref{ap:lemma2}. From \eqref{ap:lemma1}, we should then have:
\begin{subequations}\label{ap:lemma3}
\begin{align}
   \mathbb{P}_{\mathcal{A} \mathcal{B} \mathcal{E}} \quad\quad &\xmapsto{\text{LOSR}_{\mathcal{A} \mathcal{B}}} \quad \mathbb{P}_{\mathcal{A} \mathcal{B} \mathcal{E}}^{\text{iso}(\alpha)} = \alpha \text{PR}_{\mathcal{A}\mathcal{B}}\otimes L_{\mathcal{E}} +(1-\alpha)\mathbb{P}_{\mathcal{A} \mathcal{B}\mathcal{E}}',\\\nonumber \\\beta(\mathcal{A},\mathcal{B})_{\mathbb{P}_{\mathcal{A} \mathcal{B} \mathcal{E}}} &\quad\:=\quad \quad\beta(\mathcal{A},\mathcal{B})_{\mathbb{P}_{\mathcal{A} \mathcal{B} \mathcal{E}}^{\text{iso}(\alpha)}} = \frac{1+\alpha}{2},\label{ap:lemma3b}
\end{align}
\end{subequations}
where $L_{\mathcal{E}}$ denotes some local distribution for the party $\mathcal{E}$, and $ \mathbb{P}_{\mathcal{A} \mathcal{B}\mathcal{E}}'$ is any tripartite distribution addressing \eqref{ap:lemma2}, \ie, $\sum_{\mathcal{E}} \mathbb{P}_{\mathcal{A} \mathcal{B}\mathcal{E}}' = W_{\mathcal{A} \mathcal{B}}$, such that $\beta(\mathcal{A},\mathcal{B})_{\mathbb{P}_{\mathcal{A} \mathcal{B} \mathcal{E}}'}=1/2$.
Notice that $ \mathbb{P}_{\mathcal{A} \mathcal{B}\mathcal{E}}'$ might, in principle, be a non-local correlation. Importantly, the product structure of the correlation $\text{PR}_{\mathcal{A}\mathcal{B}}\otimes L_{\mathcal{E}}$ follows from the respective no-signaling conditions: Specifically, the no-signaling condition forces the tripartite distribution to \emph{factorize} whenever any two of the three parties share a PR-box \cite{Monogamy_NS}.

Naturally, the transformation described in \eqref{ap:lemma3} can also be written in terms of the $\mathcal{B}\mathcal{E}$ marginal, leaving the value of $\beta(\mathcal{B},\mathcal{E})$ unchanged. More importantly, since depolarization involves only bit-flip operations (see \cite{GeneralProperties}), the CHSH value for any marginal remains invariant under the transformation in \eqref{ap:lemma3}. This invariance becomes particularly clear when noting that $\beta(\mathcal{A},\mathcal{B})_{\mathbb{P}_{\mathcal{A} \mathcal{B} \mathcal{E}}}$ depends only on the parameter $\alpha$ from Eq. \eqref{ap:lemma3b}. Therefore, applying two successive depolarizations—$\text{LOSR}_{\mathcal{A} \mathcal{B}}$ followed by $\text{LOSR}_{\mathcal{B} \mathcal{E}}$—yields:
\begin{subequations}\label{ap:lemma4}
\begin{align}
   \mathbb{P}_{\mathcal{A} \mathcal{B} \mathcal{E}}^{\text{iso}(\alpha)} \quad\quad &\xmapsto{\text{LOSR}_{\mathcal{B} \mathcal{E}}} \quad \mathbb{P}_{\mathcal{A} \mathcal{B} \mathcal{E}}^{\text{iso}'(\alpha, \gamma)} = \gamma \text{PR}_{\mathcal{B}\mathcal{E}}\otimes L_{\mathcal{A}}' +(1-\gamma)\mathbb{P}_{\mathcal{A} \mathcal{B}\mathcal{E}}''^{(\alpha )},\label{ap:lemma4a}\\ \nonumber\\\beta(\mathcal{B},\mathcal{E})_{\mathbb{P}_{\mathcal{A} \mathcal{B} \mathcal{E}}^{\text{iso}(\alpha)}} &\quad\:=\quad \quad\beta(\mathcal{B},\mathcal{E})_{\mathbb{P}_{\mathcal{A} \mathcal{B} \mathcal{E}}^{\text{iso}'(\alpha, \gamma)}},\\ \beta(\mathcal{A},\mathcal{B})_{\mathbb{P}_{\mathcal{A} \mathcal{B} \mathcal{E}}^{\text{iso}(\alpha)}} &\quad\:=\quad \quad\beta(\mathcal{A},\mathcal{B})_{\mathbb{P}_{\mathcal{A} \mathcal{B} \mathcal{E}}^{\text{iso}'(\alpha,\gamma)}},\label{ap:lemma4c}
\end{align}
\end{subequations}
where $\gamma \in [0,1]$, $L_{\mathcal{A}}'$ is a local distribution for $\mathcal{A}$, and $\mathbb{P}_{\mathcal{A} \mathcal{B}\mathcal{E}}''^{(\alpha)}$ is any tripartite distribution such that $\sum_{\mathcal{A}} \mathbb{P}_{\mathcal{A} \mathcal{B}\mathcal{E}}''^{(\alpha)} = W_{\mathcal{B} \mathcal{E}}$, and $\beta(\mathcal{B},\mathcal{E})_{\mathbb{P}_{\mathcal{A} \mathcal{B} \mathcal{E}}'''}=1/2$. According to \eqref{ap:lemma4c}, the transformation $\text{LOSR}_{\mathcal{B} \mathcal{E}}$ does not affect the contribution $\alpha \text{PR}_{\mathcal{A}\mathcal{B}}\otimes L_{\mathcal{E}}$ from \eqref{ap:lemma3}. At most, this component might be relabeled into another version of the PR-box, still achieving the maximal value for some relabeled CHSH inequality. Hence, we may write $\mathbb{P}_{\mathcal{A} \mathcal{B}\mathcal{E}}''^{(\alpha)}$ in \eqref{ap:lemma4a} in terms of one more parameter $\epsilon\in[0,1]$, such that:
\begin{align}\label{ap:lemma5}
    \mathbb{P}_{\mathcal{A} \mathcal{B}\mathcal{E}}''^{(\alpha,\epsilon)} = \epsilon \text{PR}_{\mathcal{A}\mathcal{B}}\otimes L_{\mathcal{E}} + (1-\epsilon) \mathbb{P}_{\mathcal{A} \mathcal{B}\mathcal{E}}'''.
\end{align}
$\mathbb{P}_{\mathcal{A} \mathcal{B}\mathcal{E}}'''$ should address the constraints of $ \mathbb{P}_{\mathcal{A} \mathcal{B}\mathcal{E}}''^{(\alpha,\epsilon)}$ and $ \mathbb{P}_{\mathcal{A} \mathcal{B}\mathcal{E}}'$ in \eqref{ap:lemma3} and \eqref{ap:lemma4}, respectively. \ie, with \eqref{ap:lemma5} in \eqref{ap:lemma4a} we have,
\begin{align}\label{ap:lemma6}
    \mathbb{P}_{\mathcal{A} \mathcal{B} \mathcal{E}}^{\text{iso}' (\alpha, \gamma)} = \alpha \text{PR}_{\mathcal{A}\mathcal{B}}\otimes L_{\mathcal{E}}+ \gamma \text{PR}_{\mathcal{B}\mathcal{E}}\otimes L_{\mathcal{A}}' +(1- \alpha - \gamma) \mathbb{P}_{\mathcal{A} \mathcal{B}\mathcal{E}}'''.
\end{align}
and $\mathbb{P}_{\mathcal{A} \mathcal{B}\mathcal{E}}'''$ should respect,
\begin{subequations}\label{ap:lemma7}
\begin{align}
    \sum_{\mathcal{E}}&\left\{ \gamma \text{PR}_{\mathcal{B}\mathcal{E}}\otimes L_{\mathcal{A}}' + (1-\gamma) \mathbb{P}_{\mathcal{A} \mathcal{B}\mathcal{E}}''' \right\} = W_{\mathcal{A}\mathcal{B}},\\
    \sum_{\mathcal{A}}&\left\{ \epsilon \text{PR}_{\mathcal{A}\mathcal{B}}\otimes L_{\mathcal{E}} + (1-\epsilon) \mathbb{P}_{\mathcal{A} \mathcal{B}\mathcal{E}}''' \right\} = W_{\mathcal{B}\mathcal{E}}.
\end{align}
\end{subequations}
Notice that in \eqref{ap:lemma6} we considered the fact $\alpha = (1-\gamma)\epsilon$. At this point, it follows as a corollary from the \textit{depolarization} procedure \eqref{ap:lemma1} that tripartite correlations of binary inputs and binary outputs Bell scenarios can be transformed to the distribution \eqref{ap:lemma6}, without changing the CHSH values, $\beta(\mathcal{A},\mathcal{B})$ and $\beta(\mathcal{B},\mathcal{E})$. In particular, in each of these slices, the value of the CHSH expressions are determined exclusively by the coefficients ${(\alpha,\gamma)}$ such that,
\begin{align} \label{CHSHAB}
    \beta(\mathcal{A},\mathcal{B})_{\mathbb{P}_{\mathcal{A} \mathcal{B} \mathcal{E}}^{\text{iso}' (\alpha, \gamma)}}=\frac{1+\alpha}{2}; \\ \label{CHSHAE}
    \beta(\mathcal{B},\mathcal{E})_{\mathbb{P}_{\mathcal{A} \mathcal{B} \mathcal{E}}^{\text{iso}'(\alpha, \gamma)}}=\frac{1+\gamma}{2}.
\end{align}

We are now prepared to address the central problem: Finding the maximum value of $\beta(\mathcal{B},\mathcal{E})$ given a fixed value of $\beta(\mathcal{A},\mathcal{B})$, under the constraints imposed by the information-theoretic criteria \eqref{eq:firstICnoisy}, \eqref{eq:recentIC}, and \eqref{eq:tripartite_IC} associated with the $\mathcal{IC}$ principle. This can be framed as the following optimization problem:
\begin{align}\label{ap:optmization}
    \text{max}& \quad \beta(\mathcal{B},\mathcal{E})_{\mathbb{P}_{\mathcal{A} \mathcal{B}\mathcal{E}}}\nonumber\\
    \text{subj to}& \quad  \mathcal{I}_{\mathbb{P}_{\mathcal{A} \mathcal{B}\mathcal{E}}}^q \le \mathcal{C}^q\\
    & \quad \beta(\mathcal{A},\mathcal{B}) = p\nonumber
\end{align}
for a fixed value $p\in[0.5 , 1]$. The \ic~constraints in Eqs. \eqref{eq:firstICnoisy}, \eqref{eq:recentIC}, \eqref{eq:multipartite_IC}, and \eqref{eq:tripartite_IC} are generically expressed as $\mathcal{I}_{\mathbb{P}_{\mathcal{A} \mathcal{B}\mathcal{E}}}^q \le \mathcal{C}^q$, where $\mathcal{I}_{\mathbb{P}_{\mathcal{A} \mathcal{B}\mathcal{E}}}^q$ is the \ic~functional—i.e., a sum of mutual information terms depending on $\mathbb{P}_{\mathcal{A} \mathcal{B}\mathcal{E}}$—and $\mathcal{C}^q$ denotes constant parameters fixed by the protocol and communication channel. It is evident the \ic~functional $\mathcal{I}$ depends on the joint probability distribution, $P$, of all variables involved in the scenario, however, since the protocol is considered fixed for \eqref{ap:optmization}, we may write $\mathcal{I}_P^q = \mathcal{I}_{\mathbb{P}_{\mathcal{A} \mathcal{B}\mathcal{E}}}^q$.

Suppose there is an optimal tripartite distribution, $\mathbb{P}_{\mathcal{A} \mathcal{B}\mathcal{E}}^{*}$, solving the optimization problem \eqref{ap:optmization}. As we demonstrated, this distribution can be transformed via LOSR into $\mathbb{P}_{\mathcal{A} \mathcal{B} \mathcal{E}}^{\text{iso}' (\alpha, \gamma)}$, keeping the CHSH values, $\beta(\mathcal{A},\mathcal{B})$ and $\beta(\mathcal{B},\mathcal{E})$. In parallel, as LOSR constitutes a local post-processing operation, which, according to the data processing inequality \eqref{eq:dataprocessing} yields:
\begin{align}\label{ap:dataprocessing_lemma}
    \mathcal{I}_{\mathbb{P}_{\mathcal{A} \mathcal{B} \mathcal{E}}^{\text{iso}' (\alpha, \gamma)}}^q \le \mathcal{I}_{\mathbb{P}_{\mathcal{A} \mathcal{B}\mathcal{E}}^{*}}^q.
\end{align}
As the upper bounds, $\mathcal{C}^q$, have no dependence on the correlation, $\mathbb{P}_{\mathcal{A} \mathcal{B}\mathcal{E}}^{*}$, the distribution resulting from the procedure $\mathbb{P}_{\mathcal{A} \mathcal{B}\mathcal{E}}^{*} \xmapsto{\text{LOSR}}\mathbb{P}_{\mathcal{A} \mathcal{B} \mathcal{E}}^{\text{iso}' (\alpha, \gamma)}$ cannot violate any of the information-theoretic constraints from $\mathcal{IC}$. Consequently, the optimal CHSH values as permitted by \ic~are always achieved by $\mathbb{P}_{\mathcal{A} \mathcal{B} \mathcal{E}}^{\text{iso}' (\alpha, \gamma)}$. In other words, $\mathbb{P}_{\mathcal{A} \mathcal{B} \mathcal{E}}^{\text{iso}' (\alpha, \gamma)}$ encompasses all optimal values of the problem \eqref{ap:optmization}, \ie,
 \begin{align}
     \beta(\mathcal{B},\mathcal{E})_{\mathbb{P}_{\mathcal{A} \mathcal{B}\mathcal{E}}^{*}} \in \left\{\beta(\mathcal{B},\mathcal{E})_{\mathbb{P}_{\mathcal{A} \mathcal{B} \mathcal{E}}^{\text{iso}' (\alpha, \gamma)}} \right\}_{\alpha, \gamma}.
 \end{align}

We now turn to identifying the specific distributions  $L_{\mathcal{A}}'^*, L_{\mathcal{E}}^*, \mathbb{P}_{\mathcal{A} \mathcal{B}\mathcal{E}}'''^*$ that solve the optimization problem in Eq.\eqref{ap:optmization}. Let us first focus in $\mathbb{P}_{\mathcal{A} \mathcal{B}\mathcal{E}}'''^*$. In this case, we note that maximally uncorrelated distribution $W_{\mathcal{A}\mathcal{B}\mathcal{E}}$ ($p_{W_{\mathcal{A}\mathcal{B}\mathcal{E}}}(a,b,e|x,y,z) = 1/8$) provides the minimum value for the \ic~functional, \ie, $\mathcal{I}_{W_{\mathcal{A}\mathcal{B}\mathcal{E}}}^q \le \mathcal{I}_{\mathbb{P}_{\mathcal{A}\mathcal{B}\mathcal{E}}}^q$, $\forall\: \mathbb{P}_{\mathcal{A}\mathcal{B}\mathcal{E}}$. As a consequence, any other distribution, $\mathbb{P}_{\mathcal{A} \mathcal{B}\mathcal{E}}'''$, different of $W_{\mathcal{A}\mathcal{B}\mathcal{E}}$ cannot yield on smaller values for the \ic~functional than the case $\mathbb{P}_{\mathcal{A} \mathcal{B}\mathcal{E}}''' = W_{\mathcal{A}\mathcal{B}\mathcal{E}}$. This can be easily seen by invoking the \emph{convexity of mutual information} on probability distributions, which considering \eqref{ap:lemma6} gives,
\begin{subequations}\label{ap:lemma8}
\begin{align}
    &\mathcal{I}_{\mathbb{P}_{\mathcal{A} \mathcal{B} \mathcal{E}}^{\text{iso}' (\alpha, \gamma)}}^q \;\;\le \alpha\mathcal{I}_{\text{PR}_{\mathcal{A}\mathcal{B}}\otimes L_{\mathcal{E}}}^q + \gamma\mathcal{I}_{\text{PR}_{\mathcal{B}\mathcal{E}}\otimes L_{\mathcal{A}}'}^q + (1-\alpha - \gamma)\mathcal{I}_{ \mathbb{P}_{\mathcal{A} \mathcal{B}\mathcal{E}}'''}^q,\label{ap:lemma8a}\\
    &\mathcal{I}_{\mathbb{P}_{\mathcal{A} \mathcal{B} \mathcal{E}}^{\text{iso}' (\alpha, \gamma)W} }^q \le\alpha\mathcal{I}_{\text{PR}_{\mathcal{A}\mathcal{B}}\otimes L_{\mathcal{E}}}^q + \gamma\mathcal{I}_{\text{PR}_{\mathcal{B}\mathcal{E}}\otimes L_{\mathcal{A}}'}^q + (1-\alpha - \gamma)\mathcal{I}_{ W_{\mathcal{A} \mathcal{B}\mathcal{E}}}^q,\label{ap:lemma8b}
\end{align}
\end{subequations}
where $\mathbb{P}_{\mathcal{A} \mathcal{B} \mathcal{E}}^{\text{iso}' (\alpha, \gamma)W}$ denotes the particular case of $\mathbb{P}_{\mathcal{A} \mathcal{B} \mathcal{E}}^{\text{iso}' (\alpha, \gamma)}$ when $\mathbb{P}_{\mathcal{A} \mathcal{B}\mathcal{E}}''' = W_{\mathcal{A}\mathcal{B}\mathcal{E}}$ in \eqref{ap:lemma6}. Subtracting both relations in \eqref{ap:lemma8} and considering $\mathcal{I}_{W_{\mathcal{A}\mathcal{B}\mathcal{E}}}^q \le \mathcal{I}_{\mathbb{P}_{\mathcal{A}\mathcal{B}\mathcal{E}}'''}^q$, we have 
\begin{align}\label{ap:lemma9}
    \mathcal{I}_{\mathbb{P}_{\mathcal{A} \mathcal{B} \mathcal{E}}^{\text{iso}' (\alpha, \gamma)W}}^q \le \mathcal{I}_{\mathbb{P}_{\mathcal{A} \mathcal{B} \mathcal{E}}^{\text{iso}' (\alpha, \gamma)}}^q.
\end{align}
In parallel, from \eqref{ap:lemma7}, it follows that the distribution $\mathbb{P}_{\mathcal{A} \mathcal{B}\mathcal{E}}'''$ does not contribute to the CHSH values; that is, $\beta(\mathcal{A},\mathcal{B})_{\mathbb{P}_{\mathcal{A} \mathcal{B}\mathcal{E}}'''} = \beta(\mathcal{B},\mathcal{E})_{\mathbb{P}_{\mathcal{A} \mathcal{B}\mathcal{E}}'''} = 1/2$. Combining this fact with \eqref{ap:lemma9}, we conclude that any distribution $\mathbb{P}_{\mathcal{A} \mathcal{B}\mathcal{E}}'''$, different from $W_{\mathcal{A}\mathcal{B}\mathcal{E}}$, performs at most as good as the case $\mathbb{P}_{\mathcal{A} \mathcal{B}\mathcal{E}}''' = W_{\mathcal{A}\mathcal{B}\mathcal{E}}$. This is because \eqref{ap:lemma9} ensures that, in general, any such $\mathbb{P}_{\mathcal{A} \mathcal{B}\mathcal{E}}''' $ brings the value of the \ic~functional closer to the upper bound $\mathcal{C}^q$—regardless of $\alpha$ and $\gamma$—without contributing for the increasing of $\beta(\mathcal{B},\mathcal{E})$. Consequently, any value $\beta(\mathcal{B},\mathcal{E})_{\mathbb{P}_{\mathcal{A} \mathcal{B} \mathcal{E}}^{\text{iso}' (\alpha, \gamma)}}$, such that $\beta(\mathcal{B},\mathcal{E})_{\mathbb{P}_{\mathcal{A} \mathcal{B} \mathcal{E}}^{\text{iso}' (\alpha, \gamma)}} > \max \{\beta(\mathcal{B},\mathcal{E})_{\mathbb{P}_{\mathcal{A} \mathcal{B} \mathcal{E}}^{\text{iso}' (\alpha, \gamma)W}}\}$, will result in a violation of \ic, as ensured by \eqref{ap:lemma9}. Therefore, it is sufficient to take the optimal distribution $\mathbb{P}_{\mathcal{A} \mathcal{B}\mathcal{E}}^{*}$ as the following form:
 \begin{align}
     \mathbb{P}_{\mathcal{A} \mathcal{B}\mathcal{E}}^{*(\alpha,\gamma)} = \mathbb{P}_{\mathcal{A} \mathcal{B} \mathcal{E}}^{\text{iso}' (\alpha, \gamma)W} =  \alpha \text{PR}_{\mathcal{A}\mathcal{B}}\otimes L_{\mathcal{E}}^*+ \gamma \text{PR}_{\mathcal{B}\mathcal{E}}\otimes L_{\mathcal{A}}'^* +(1- \alpha - \gamma) W_{\mathcal{A}\mathcal{B}\mathcal{E}},
 \end{align}
which is a family of two-parameter slices of tripartite no-signaling polytope characterized by the local distributions $L_{\mathcal{A}}'^*,L_{\mathcal{E}}^*$.

Next, we show that in order to derive monogamy relations of the form \eqref{eq:monogamy_T}—both for the bipartite cases \eqref{eq:firstICnoisy}, \eqref{eq:recentIC} and the multipartite criteria \eqref{eq:multipartite_IC}, \eqref{eq:tripartite_IC} associated with $\mathcal{IC}$—the optimal choice for the local distributions is \textit{white noise}: specifically, $L_{\mathcal{A}}'^* = W_{\mathcal{A}}$ and $L_{\mathcal{E}}^* = W_{\mathcal{E}}$, where $p_{\text{W}_{\mathcal{A}}}(a|x) = p_{\text{W}_{\mathcal{E}}}(e|z) = 1/2$. To establish this claim, consider the distribution $\bar{\mathbb{P}}^{*(\alpha,\gamma)}_{\mathcal{A}\mathcal{B}\mathcal{E}}$, obtained from $\mathbb{P}^{*(\alpha,\gamma)}_{\mathcal{A}\mathcal{B}\mathcal{E}}$ by flipping all outputs. This leads to:
\begin{align}
    p_{\bar{\mathbb{P}}^{*(\alpha,\gamma)}_{\mathcal{A}\mathcal{B}\mathcal{E}}}(a,b,e|x,y,z) = \alpha p_{\text{PR}_{\mathcal{A}\mathcal{B}}\otimes \bar{L}_{\mathcal{E}}^*}(a,b,e|x,y,z)+\gamma p_{\text{PR}_{\mathcal{B}\mathcal{E}}\otimes \bar{L}_{\mathcal{A}}^*}(&a,b,e|x,y,z)\\&+(1-\alpha-\gamma)p_{W_{\mathcal{A}\mathcal{B}\mathcal{E}}}(a,b,e|x,y,z), \nonumber
\end{align}
where, $p_{\bar{L}_{\mathcal{E}}^*}(e|z)=p_{{L}_{\mathcal{E}}^*}(e\oplus 1|z)$ and $p_{\bar{L}_{\mathcal{A}}^*}(a|x)=p_{{L}_{\mathcal{A}}^*}(a\oplus 1|x)$. As flipping the outputs is a simple local post-processing, similarly to \eqref{ap:dataprocessing_lemma}, data processing inequality \eqref{eq:dataprocessing} ensures that the \ic~functionals on bipartite \eqref{eq:firstICnoisy},\eqref{eq:recentIC} and multipartite criteria \eqref{eq:multipartite_IC},\eqref{eq:tripartite_IC} should only decrease, \ie:
\begin{align}\label{ap:dataprocessing_lemma1}
\mathcal{I}_{\bar{\mathbb{P}}_{\mathcal{A}\mathcal{B}\mathcal{E}}^{*(\alpha,\gamma)}}^q \le \mathcal{I}_{\mathbb{P}_{\mathcal{A}\mathcal{B}\mathcal{E}}^{*(\alpha,\gamma)}}^q.
\end{align}
Moreover, as previously mentioned for the LOSR procedures, the upper bound, $\mathcal{C}^q$, remains unaltered. Hence, given a distribution, ${\mathbb{P}}_{\mathcal{A}\mathcal{B}\mathcal{E}}^{*(\alpha,\gamma)}$, which satisfies a \ic~criterion of interest, $\mathcal{I}_{\mathbb{P}_{\mathcal{A}\mathcal{B}\mathcal{E}}^{*(\alpha,\gamma)}}^q \le \mathcal{C}^q$, then the flipped-outcomes distribution, $\bar{\mathbb{P}}_{\mathcal{A}\mathcal{B}\mathcal{E}}^{*(\alpha,\gamma)}$, also satisfies the same criterion $q$. Note, that the value of the CHSH expression, $\beta(\mathcal{A},\mathcal{B})$ and $\beta(\mathcal{B},\mathcal{E})$ remain unaltered when all outcomes are \emph{simultaneously} flipped, as they only depend on the coefficients $(\alpha,\gamma)$ \eqref{CHSHAB} \eqref{CHSHAE}.  

Now, let us consider another tripartite distribution, $\tilde{\mathbb{P}}^{*(\alpha,\gamma)}_{\mathcal{A}\mathcal{B}\mathcal{E}}$, obtained by mixing equal proportions of the original tripartite distribution ${\mathbb{P}}^{*(\alpha,\gamma)}_{\mathcal{A}\mathcal{B}\mathcal{E}}$ and the flipped-outcomes distribution $\bar{\mathbb{P}}^{*(\alpha,\gamma)}_{\mathcal{A}\mathcal{B}\mathcal{E}}$, i.e., $\tilde{\mathbb{P}}^{*(\alpha,\gamma)}_{\mathcal{A}\mathcal{B}\mathcal{E}}=\frac{1}{2}({\mathbb{P}}^{*(\alpha,\gamma)}_{\mathcal{A}\mathcal{B}\mathcal{E}}+ \bar{\mathbb{P}}^{*(\alpha,\gamma)}_{\mathcal{A}\mathcal{B}\mathcal{E}})$, such that,
\begin{equation}\label{ap:lemma10}p_{\tilde{\mathbb{P}}^{*(\alpha,\gamma)}_{\mathcal{A}\mathcal{B}\mathcal{E}}}(a,b,e|x,y,z)=\frac{1}{2}p_{\mathbb{P}^{*(\alpha,\gamma)}_{\mathcal{A}\mathcal{B}\mathcal{E}}}(a,b,e|x,y,z)+\frac{1}{2}p_{\bar{\mathbb{P}}^{*(\alpha,\gamma)}_{\mathcal{A}\mathcal{B}\mathcal{E}}}(a,b,e|x,y,z),
\end{equation}
and specifically,
\begin{align} \label{desiredForm} p_{\tilde{\mathbb{P}}^{*(\alpha,\gamma)}_{\mathcal{A}\mathcal{B}\mathcal{E}}}(a,b,e|x,y,z) &= \alpha p_{\text{PR}_{\mathcal{A}\mathcal{B}}\otimes W_{\mathcal{E}}}(a,b,e|x,y,z)+\gamma p_{\text{PR}_{\mathcal{B}\mathcal{E}}\otimes W_{\mathcal{A}}}(a,b,e|x,y,z)\\& \quad\quad\quad\quad\quad\quad\quad\quad\quad\quad\quad\quad\quad\quad\quad\quad\quad\quad\quad\quad\quad +(1-\alpha-\gamma)p_{W_{\mathcal{A}\mathcal{B}\mathcal{E}}}(a,b,e|x,y,z),\nonumber
\end{align}
where we have used the facts, $\frac{1}{2}(L_{\mathcal{B}}^*+\bar{L}_{\mathcal{B}}^*)=W_{\mathcal{B}}$, $\frac{1}{2}(L_{\mathcal{E}}^*+\bar{L}_{\mathcal{E}}^*)=W_{\mathcal{E}}$, $\frac{1}{2}\left(W_{\mathcal{A}\mathcal{B}\mathcal{E}}+\bar{W}_{\mathcal{A}\mathcal{B}\mathcal{E}}\right) = W_{\mathcal{A}\mathcal{B}\mathcal{E}}$. Notice that the values of the CHSH expression, $\beta(\mathcal{A},\mathcal{B})$ and $\beta(\mathcal{B},\mathcal{E})$ still remain unaltered as they only depend on the coefficients $(\alpha,\gamma)$ in \eqref{CHSHAB} and \eqref{CHSHAE}. At this point, we invoke \emph{convexity of mutual information} once more, which allows to write for the \ic~functional considering \eqref{ap:lemma10} :
\begin{align}
    \mathcal{I}_{\tilde{\mathbb{P}}^{*(\alpha,\gamma)}_{\mathcal{A}\mathcal{B}\mathcal{E}}}^q \le \frac{1}{2} \mathcal{I}_{\mathbb{P}^{*(\alpha,\gamma)}_{\mathcal{A}\mathcal{B}\mathcal{E}}}^q + \frac{1}{2}\mathcal{I}_{\bar{\mathbb{P}}^{*(\alpha,\gamma)}_{\mathcal{A}\mathcal{B}\mathcal{E}}}^q.
\end{align}
Thus, if the distribution ${\mathbb{P}}_{\mathcal{A}\mathcal{B}\mathcal{E}}^{*(\alpha,\gamma)}$ satisfies a \ic~criterion of interest, $\mathcal{I}_{\mathbb{P}_{\mathcal{A}\mathcal{B}\mathcal{E}}^{*(\alpha,\gamma)}}^q \le \mathcal{C}^q$, the distribution, $\tilde{\mathbb{P}}^{*(\alpha,\gamma)}_{\mathcal{A}\mathcal{B}\mathcal{E}}$, obtained from \eqref{ap:lemma10} will necessarily satify it, having the same value of the CHSH expressions \eqref{CHSHAB} and \eqref{CHSHAE} as ${\mathbb{P}}_{\mathcal{A}\mathcal{B}\mathcal{E}}^{*(\alpha,\gamma)}$. Consequently, any optimal distribution, ${\mathbb{P}}_{\mathcal{A}\mathcal{B}\mathcal{E}}^{*(\alpha, \gamma)}$, solving the problem \eqref{ap:optmization}, characterized by the optimal local distributions $L_{\mathcal{A}}'^*$ and $L_{\mathcal{E}}^*$, will produce the same CHSH values as $\tilde{\mathbb{P}}^{*(\alpha,\gamma)}_{\mathcal{A}\mathcal{B}\mathcal{E}}$, while $\tilde{\mathbb{P}}^{* (\alpha,\gamma)}_{\mathcal{A}\mathcal{B}\mathcal{E}}$ also satisfies \ic. Thus, in order to find the optimal CHSH values as permitted by \ic~criteria in the problem \eqref{ap:optmization} it is enough to analyze the slice \eqref{desiredForm}, which assumes the form presented in the main text:
\begin{align}
p_{{\mathbb{P}}_{\mathcal{A}\mathcal{B}\mathcal{E}}^{*(\alpha, \gamma)}}(a,b,e|x,y,z) = p_{\tilde{\mathbb{P}}^{(\alpha,\gamma)}_{\mathcal{A}\mathcal{B}\mathcal{E}}}(a,b,e|x,y,z) = \alpha \frac{1}{4}\delta_{a\oplus b, xy} + \gamma \frac{1}{4}\delta_{b\oplus e, yz}  +(1-\alpha -\gamma)1/8.
\end{align}
This completes the proof, and establishes the optimality of the slice \eqref{eq:slice}.




\end{document}